\documentclass{PoS}

\title{Transport Model Studies of the Baryon-Rich Quark-Gluon
Plasma formed in Heavy Ion Collisions}

\ShortTitle{Partonic effects}

\author{Lie-Wen Chen\\
        Department of Physics, Shanghai Jiao Tong University,
        Shanghai 200240, China\\
        E-mail: \email{lwchen@sjtu.edu.cn}}

\author{\speaker{Che Ming Ko}\\
        Cyclotron Institute and Physics Department, Texas A$\&$M
        University, College Station, Texas 77843-3366\\
        E-mail: \email{ko@comp.tamu.edu}}

\author{Wei Liu\footnote{Present address: Department of Finance, Texas
A\&M University, College Station, Texas 77843-4218}\\
        Cyclotron Institute and Physics Department, Texas A$\&$M
        University, College Station, Texas 77843-3366\\
        E-mail: \email{weiliu@comp.tamu.edu}}

\author{Ben-Wei Zhang\\
        Institute of Particle Physics, Huazhong Normal University, Wuhan 430079,
        China\\
        E-mail: \email{bwzhang@iopp.ccnu.edu.cn}}

\abstract{Heavy ion collisions in the low energy run at Relativistic
Heavy Ion Collider (RHIC) and future Facility for Antiproton and Ion
Research (FAIR) in Germany are expected to produce a quark-gluon
plasma that has a finite baryon chemical potential, allowing thus
the possibility to study the location of the critical endpoint in
the QCD phase diagram. In this talk, using a multiphase transport
model, that includes interactions in both initial partonic and final
hadronic matters and the transition between these two phases of
matter, we discuss the effects of partonic interactions on
observables such as the elliptic flow that have played essential
roles in studying the properties of the net baryon free quark-gluon
plasma produced in heavy ion collisions at higher energies at RHIC.
Also, we study the effect of density fluctuations due to a
first-order transition between the quark-gluon plasma and hadronic
matter on fluctuations of hadron mean transverse momentum and
produced deuteron number as well as on two-pion correlations.  We
further discuss the possibility of studying the mechanism of charm
energy loss in the baryon-rich quark-gluon plasma and the properties
of phi mesons in hot-dense matter produced in these collisions.}

\FullConference{5th International Workshop on Critical Point and Onset
of Deconfinement - CPOD 2009,\\
June 08 - 12 2009\\
Brookhaven National Laboratory, Long Island, New York, USA}

\begin{document}

\section{Introduction}

Many observables have been proposed as possible signatures for the
deconfined plasma of quarks and gluons that is produced during
initial stage of ultra-relativistic heavy ion collisions. These
include enhanced production of dileptons of intermediate invariant
masses \cite{shuryak} and baryons made of multi-strange quarks
\cite{rafelski}, increased emission duration
\cite{Pratt:su,bertsch}, suppressed production of charmonia
\cite{matsui}, large anisotropic flows of hadrons
\cite{Ollitrault:1992bk}, quenching of minijets with large
transverse momenta \cite{wang}, and scaling of hadron elliptic flows
according to their constituent quark content \cite{voloshin}. Most
of these observables have been studied during past many years in
experiments at the Relativistic Heavy Ion Collider (RHIC) involving
Au+Au collisions at center-of-mass energies $\sqrt{s_{NN}}=62$,
$130$, and $200$ GeV. Studying these signatures using various
theoretical models, such as the statistical model
\cite{braun,torrieri}, the hydrodynamic model
\cite{Teaney:v2,huovinen,hirano}, the transport model
\cite{Zhang:2000bd,Lin:2001cx,Lin:2004en,Molnar:2001ux,cassing,bass,sa,zhou,bravina,xu},
the quark coalescence model \cite{hwa,greco,fries,linmolnar}, and
the perturbative QCD approach \cite{gyulassy,levai}, has provided
convincing evidence that the quark-gluon plasma has indeed been
produced in these collisions. Moreover, these studies have indicated
that the quark-gluon plasma produced at RHIC is strongly interacting
with transport coefficients very different from those given by the
perturbative QCD \cite{star}.

Heavy ion collisions at energies much higher than that at RHIC will
soon be available at the Large Hadron Collider (LHC), and it is
expected that the produced quark-gluon plasma will have even higher
temperature and smaller baryon chemical potential than that produced
in heavy ion collisions at RHIC. On the other hand, a quark-gluon
plasma with finite baryon chemical potential is expected to be
produced in low energy runs at RHIC and at FAIR.  Although lattice
QCD calculations have not been able to address the properties of
quark-gluon plasma at finite baryon chemical potential, theoretical
models such as the PNJL model have shown that the smooth crossover
from the quark-gluon plasma to the hadronic matter transition
predicted by lattice gauge calculations for zero baryon chemical
potential would change to a first-order phase transition when the
baryon chemical potential becomes sufficient large
\cite{stephanov06}. Low energy runs at RHIC and FAIR thus allows the
possibility to study the location of the critical endpoint in the
QCD phase diagram at which the first-order phase transition changes
to a smooth crossover. It is therefore of great interest to make
predictions for above mentioned observables in these collisions
based on what we have learnt from heavy ion collisions at high
energy runs at RHIC. Using a multiphase transport (AMPT) model,
which has been quite useful in understanding experimental results at
RHIC, we have carried out such a study. In particular, we have made
predictions on the rapidity distributions of various hadrons, their
elliptic flows, two-pion correlation functions, and the effects of a
first-order QGP to hadronic matter phase transition in Au+Au
collisions at center-of-mass energy of $\sqrt{s_{NN}}=7$ GeV
(corresponding to a beam energy of about $E_{beam} \approx 25$
GeV/nucleon in the laboratory system with fixed target) that is
available in the low-energy run at RHIC and FAIR.

\section{A multiphase transport model}\label{ampt}

Before presenting predicted results, we review briefly the
\textrm{AMPT} model
\cite{Zhang:2000bd,Lin:2001cx,Lin:2004en,zhang,zhangli,ko,pal}. It
is a hybrid model that uses minijet partons from hard processes and
strings from soft processes in the heavy ion jet interaction
generator (\textrm{HIJING}) model \cite{Wang:1991ht} as initial
conditions for modeling heavy-ion collisions at ultra-relativistic
energies. Time evolution of resulting minijet partons, which are
largely gluons, is described by Zhang's parton cascade
(\textrm{ZPC}) model \cite{Zhang:1997ej}. At present, this model
includes only parton-parton elastic scatterings with an in-medium
cross section given by the perturbative QCD, i.e.,
\begin{eqnarray}\label{cross}
\frac {d\sigma_p}{dt}\simeq\frac{9\pi
\alpha_s^2}{2(t-\mu^2)^2},\qquad \sigma_p\simeq\frac{9\pi
\alpha_s^2}{2\mu^2} \frac {1}{1+\mu^2/s},
\end{eqnarray}
where $\alpha _{s}$ is the strong coupling constant and is taken to
have a value of 0.47, and $s$ and $t$ are the usual Mandelstam
variables for squared center-of-mass energy and four momentum
transfer, respectively. The effective screening mass $\mu $ depends
on the temperature and density of partonic matter but is taken as a
parameter in \textrm{ZPC} for fixing the magnitude and angular
distribution of the parton scattering cross section. After minijet
partons stop interacting, they are combined with their parent
strings, as in the \textrm{HIJING} model with jet quenching, to
fragment into hadrons using the Lund string fragmentation model as
implemented in the \textrm{PYTHIA} program \cite{Sjostrand:1993yb}.
The final-state hadronic scatterings are modeled by a relativistic
transport (\textrm{ART}) model \cite{Li:1995pr,lisustich}. This
default \textrm{AMPT} model has been quite successful in describing
measured rapidity distributions of charged particles, particle to
antiparticle ratios, and spectra of low transverse momentum pions
and kaons in heavy ion collisions at the Super Proton Synchrotron
(\textrm{SPS}) and \textrm{RHIC} \cite{Lin:2001cx}. It has also been
useful in understanding the production of $J/\psi $ \cite{zhang} and
multistrange baryons \cite{pal} in these collisions.

Since the initial energy density in ultra-relativistic heavy ion
collisions is expected to be much larger than the critical energy
density at which the hadronic matter to quark-gluon plasma
transition would occur \cite{Lin:2004en,zhang,Kharzeev:2001ph}, the
AMPT model has been extended to allow initial excited strings to
melt into partons \cite{Lin:2001zk}. In this version, hadrons that
would have been produced from the HIJING model are converted to
their valence quarks and/or antiquarks. Interactions among these
partons are again described by the \textrm{ZPC} parton cascade
model. Because inelastic scatterings are not included in the current
version of the ZPC model, only quarks and antiquarks from melted
strings are present in the partonic matter. The species independence
of the cross section used in the ZPC model compensates, however, for
the absence of gluons in the early stage.

The transition from the partonic matter to the hadronic matter in
the AMPT with string melting is achieved using a coordinate-space
quark coalescence model, i.e., two nearest quark and antiquark are
combined into mesons and three nearest quarks or antiquarks are
combined into baryons or anti-baryons that are closest to the
invariant masses of these parton combinations. This coalescence
model is somewhat different from the ones that are based on the
overlap of hadron quark wave functions with the quark distribution
functions in the partonic matter and used extensively for studying
the production of hadrons with intermediate transverse momenta
\cite{hwa,greco,fries}. The final-state scatterings of produced
hadrons from quark coalescence are again described by the
\textrm{ART} model.

Using parton scattering cross sections of $6$-$10$ \textrm{mb}, the
\textrm{AMPT} model with string melting is able to reproduce the
centrality and transverse momentum (below $2$ \textrm{GeV}$/c$)
dependence of hadron elliptic flows \cite{Lin:2001zk} and
higher-order anisotropic flows \cite{Chen:2004dv} as well as the
pion interferometry \cite{LinHBT02} measured in Au+Au collisions at
$\sqrt{s_{NN}}=130$ \textrm{GeV} at \textrm{RHIC}
\cite{Ackermann:2000tr,adams,STARhbt01}. It has also been used to
study the kaon interferometry \cite{lin} in these collisions as well
as many other observables at $\sqrt{s_{NN}}=200$ GeV, such as the
pseudorapidity \cite{Chen:2004vh}, system size
\cite{chenko1,chenko2}, and flavor \cite{binzhang,chenphi}
dependence of anisotropic flows.

\section{Trajectories of central heavy ion collisions in the QCD
phase diagram}

\begin{figure}[ht]
\centerline{
\includegraphics[width=3in,height=3in,angle=0]{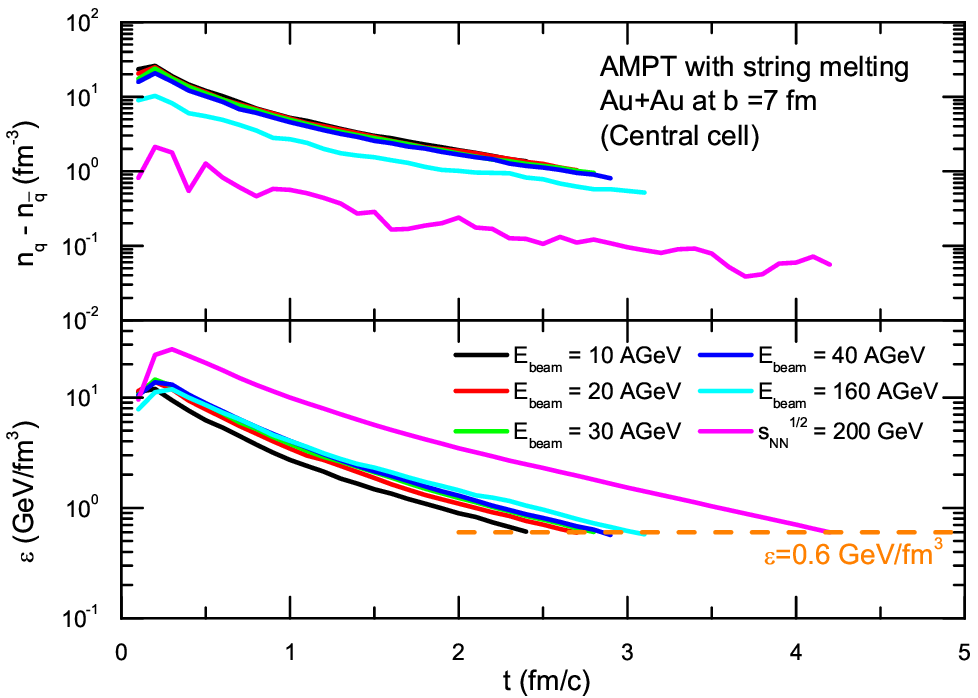}
\includegraphics[width=3in,height=3in,angle=0]{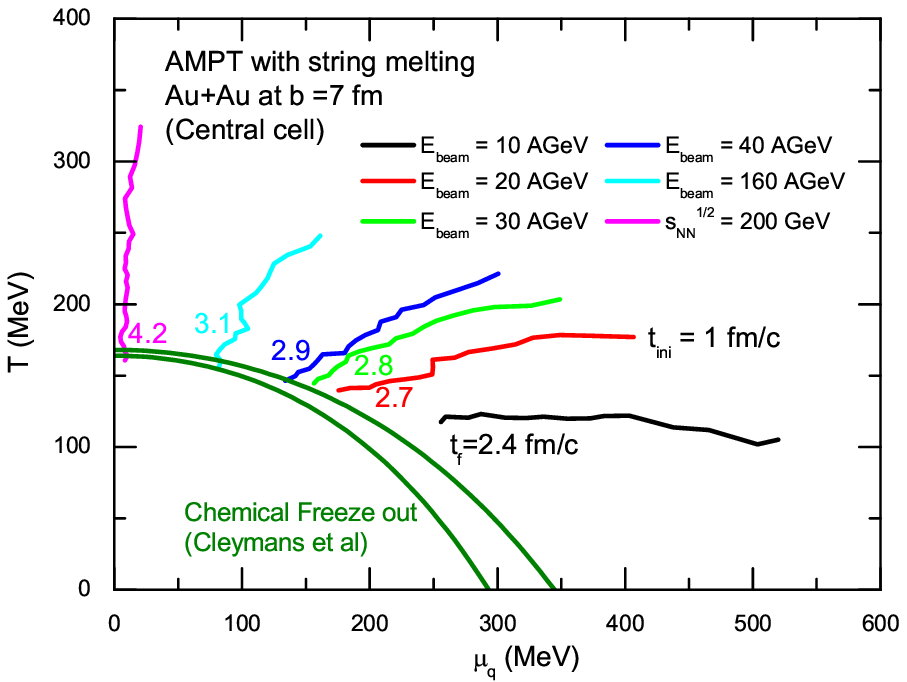}}
\caption{(Color online) Left window: Time evolution of net quark and
energy densities. Right window: Trajectories of central heavy ion
collisions in the QCD phase diagram of temperature and net quark
chemical potential. } \label{phase}
\end{figure}

Using the AMPT model with string melting, we have studied the time
evolution of net quark density and energy density of produced
partonic matter in relativistic heavy ion collisions. This is shown
in the left window of Fig.~\ref{phase} for the central cell in the
produced matter, which is taken to have a transverse radius of 1 fm
and a longitudinal dimension of 5\% of the time after the two nuclei
have fully overlapped in the longitudinal direction. If we assume
that these partons are in thermal equilibrium, their temperature can
then be determined. The time evolution of the temperature and baryon
chemical potential of produced partonic matter is shown in the right
window of Fig.~\ref{phase}. It is seen that the duration of the
partonic stage decreases slightly with collision energy, about 4.2
fm/$c$ at RHIC and less than 3 fm/$c$ at lower energies and at FAIR.
Although the baryon chemical potential (given by $3\mu_q$) in heavy
ion collisions at RHIC is small, it increases significantly as the
energy decreases.

\section{Rapidity distributions}

\begin{figure}[ht]
\centerline{
\includegraphics[width=3in,height=3in,angle=0]{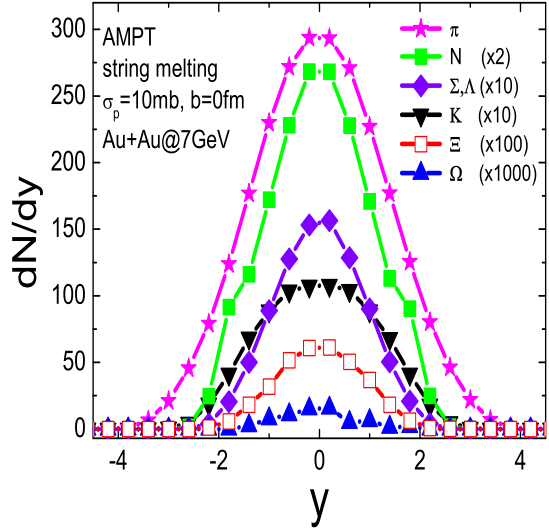}}
\caption{(Color online) Rapidity distributions of identified hadrons
in central Au+Au collisions at $\sqrt {s_{NN}}=7$ GeV from the AMPT
model with string melting.} \label{dndy}
\end{figure}

The number of particles produced in a heavy ion collision provides
valuable information on the energy density of the matter formed
during initial stage. Shown in Fig.~\ref{dndy} are the rapidity
distributions of identified hadrons in central ($b=0$ fm) Au+Au
collisions at $\sqrt{s_{NN}}=7$ GeV. The total multiplicity density
at midrapidity is about 450, which is about a factor of three lower
than that in high energy runs at RHIC. This is consistent with an
initial energy density of the central cell, which is about 10
GeV/fm$^3$, that is also about a factor of three lower than that in
high energy runs at RHIC. On the other hand, the net baryon rapidity
density at midrapidity in central heavy ion collisions at this lower
energy is about 135 and is almost half of that for pions, which is
contrary to heavy ion collisions in high energy runs at RHIC where
the net baryon rapidity density at midrapidity is less than 10. For
strange hadrons, more hyperons than kaons are produced in heavy ion
collisions at low energies as a result of the large baryon chemical
potential. This is in sharp contrast to what was observed in heavy
ion collisions in high energy runs at RHIC, where the number of
produced kaons is order of magnitude larger than that of hyperons.
The large baryon chemical potential also leads to a yield of
multistrange baryons in these collisions that is comparable to that
at RHIC.

\section{Anisotroic flows}

\begin{figure}[ht]
\vspace{0.1cm} \centerline{
\includegraphics[width=1.9in,height=1.9in,angle=0]{v2_fair_parton_new.eps}
\includegraphics[width=2in,height=2in,angle=0]{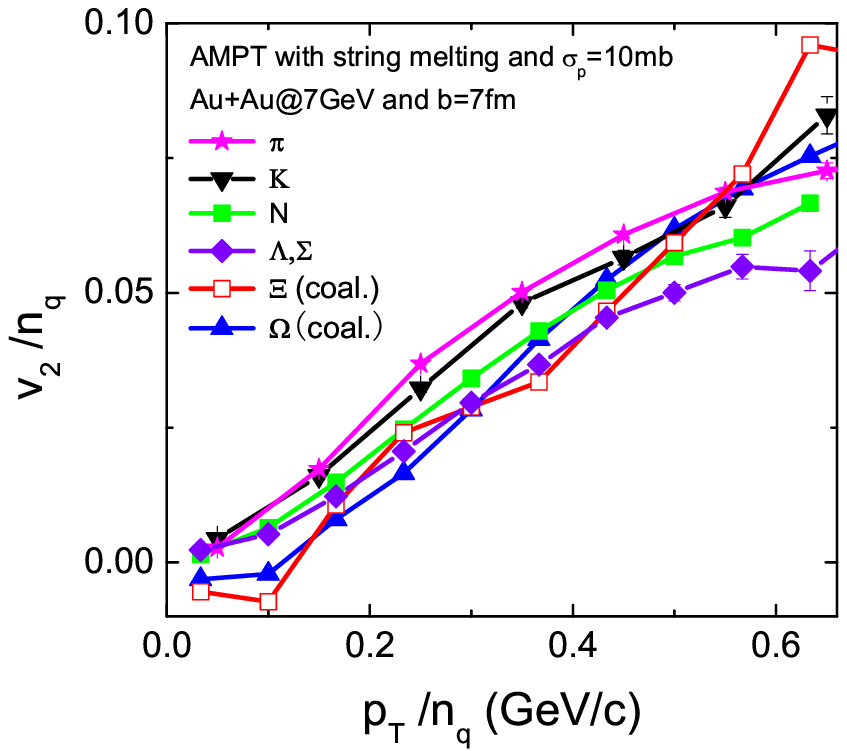}
\includegraphics[width=1.9in,height=1.9in,angle=0]{heavy_meson_v2_new.eps}}
\caption{(Color online) Differential elliptic flows of quarks (left
window), light (middle window) and heavy (right window) hadrons in
central Au+Au collisions at $\sqrt {s_{NN}}=7$ GeV.}\label{v2}
\end{figure}

In non-central heavy ion collisions, the spatial anisotropy of
initially produced matter in the transverse plane is converted to
the momentum anisotropy of produced particles. Results for Au+Au
collisions at $\sqrt{s_{NN}}=7$ GeV and impact parameter $b=7$ fm
with a parton scattering cross section of 10 mb are shown in
Fig.~\ref{v2}. As shown in the left window, the quark elliptic flows
at low transverse momentum follow the mass ordering with the charm
quark elliptic flow much smaller than that of light and strange
quarks. Their values become, however, similar at high transverse
momentum and are comparable to those in heavy ion collisions in high
energy runs at RHIC. The elliptic flows of identified hadrons, such
as the pion, kaon, nucleon, and various hyperons, are shown in the
middle window of Fig.~\ref{v2} using the scaled elliptic flow and
transverse momentum, i.e, both are divided by the number of
constituent quarks in a hadron. It is seen that the constituent
quark number scaling expected from the naive coalescence model, in
which only quarks with same momentum can coalescence into a hadron,
is not exactly satisfied as the coalescence model used in the AMPT
model for describing the hadronization of the partonic matter is
based on the coordinate-space consideration as described in Section
\ref{ampt}. In the right window of Fig.~\ref{v2}, the elliptic flows
of charmed meson and charmonium are shown. For charmed mesons, their
elliptic flow $v_{2,D}(p_T)$ at transverse momentum $p_T$ is
obtained using the quark coalescence or recombination model of
Ref.\cite{linmolnar}, i.e.,
\begin{equation}\label{heavy}
v_{2,D}(p_T)\approx v_{2,c}((m_c/m_D)p_T)+v_{2,q}((m_q/m_D)p_T).
\end{equation}
In the above, $v_{2,c}$ and $v_{2,q}$ are elliptic flows of charm
and light quarks, respectively; while $m_D$, $m_c=1.5$ GeV, and
$m_q=300$ MeV are, respectively, the masses of charmed meson, charm
quark, and light quark. Because of the much larger charm quark mass
than those of light quarks, the elliptic flows of charmed mesons are
close to that of heavy quarks. Eq.(\ref{heavy}) can be generalized
to heavy mesons with hidden charm, i.e., the charmonium $J/\psi$
that consists of a charm quark and its antiquark. Its elliptic flow
at $p_T$ is then twice that of charm quark at $p_T/2$ and is also
shown in the right window of Fig.~\ref{v2}.

The elliptic flows of heavy mesons have already been studied in high
energy runs at RHIC via measurement of their decay electrons
\cite{adler2,bielcik}. The observed large value in Au+Au collisions
at $\sqrt{s_{NN}}=200$ GeV is consistent with large elliptic flows
of heavy quarks, particular that of charmed quarks as shown in
Refs.\cite{binzhang,hees} based on the quark coalescence model.
Without heavy quark elliptic flow, resulting heavy meson elliptic
flow would be much smaller as shown in Ref.\cite{grecoko}. Studying
heavy meson elliptic flow in low energy runs at RHIC and at FAIR is
thus very useful for understanding the dynamics of heavy quarks in
the partonic matter with finite baryon chemical potential.

As in heavy ion collisions in high energy runs at RHIC, the elliptic
flows in heavy ion collisions in low-energy runs at RHIC and at FAIR
would be significantly reduced if the partonic matter is not
produced during the initial stage of the collisions.

\section{Effects of density fluctuation due to a first-order phase transition}

\begin{figure}[ht]
\centerline{
\includegraphics[width=2in,height=4in,angle=-90]{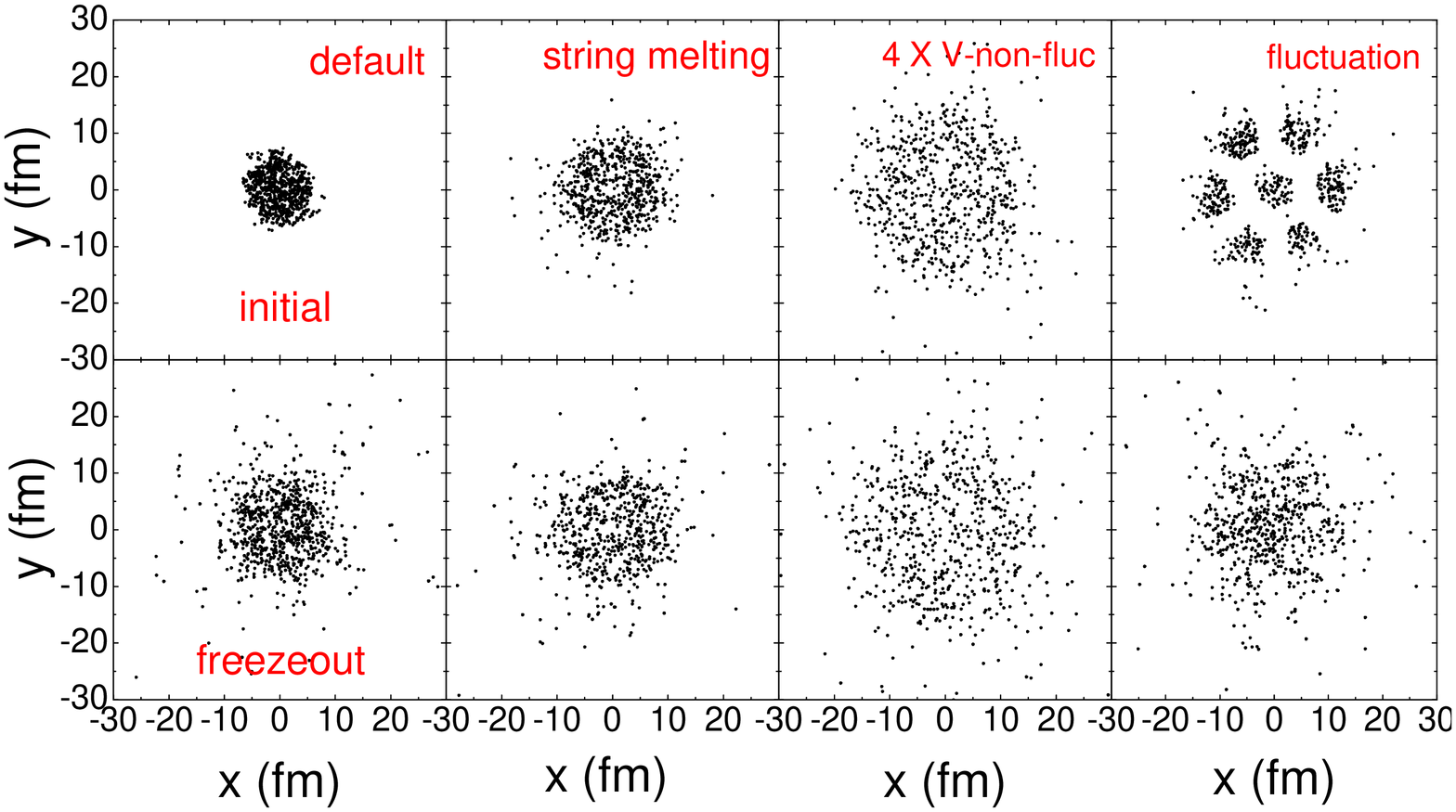}}
\vspace{0.2cm}
\centerline{
\includegraphics[width=2in,height=4in,angle=-90]{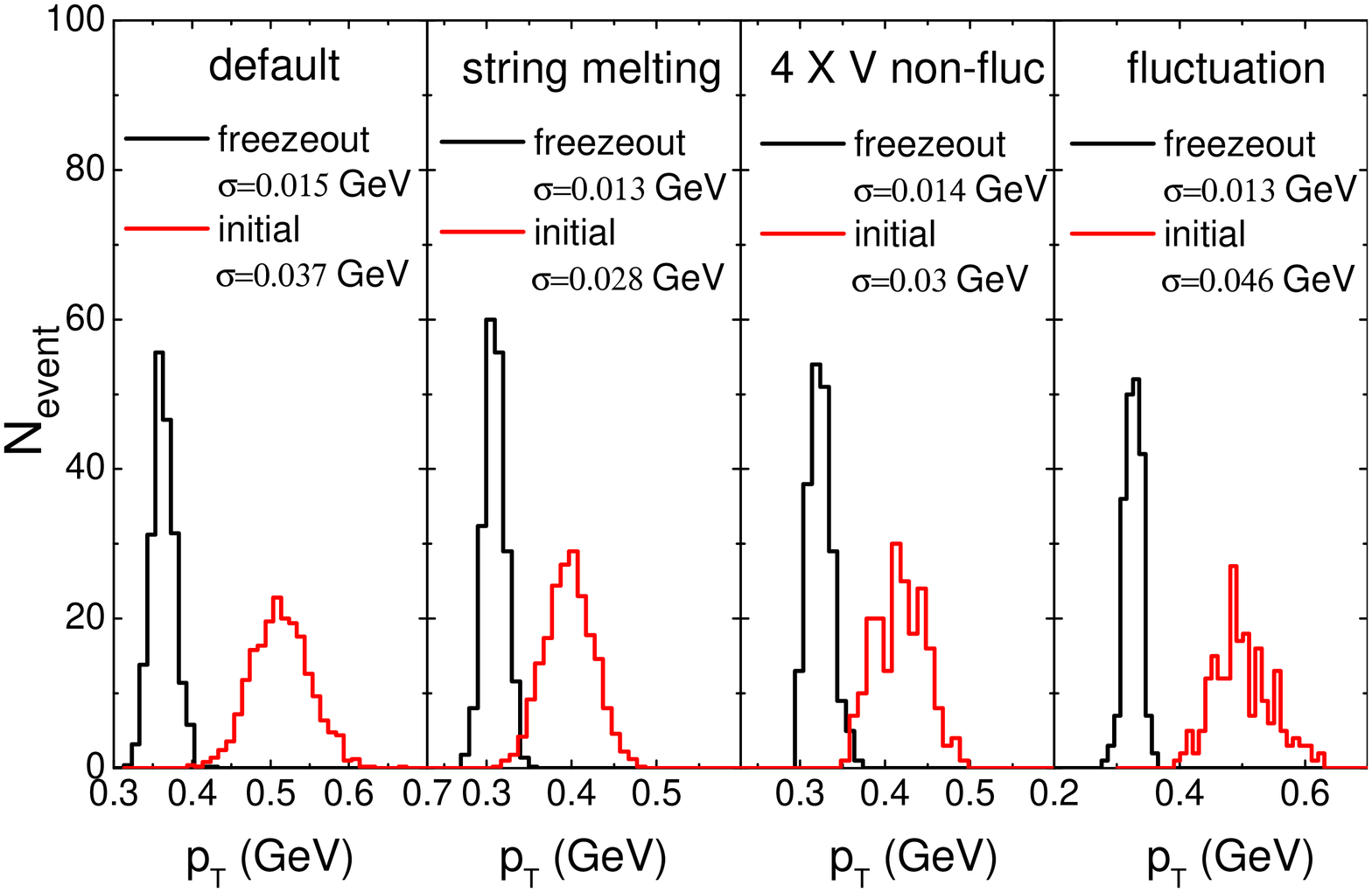}}
\vspace{0.2cm}
\centerline{
\includegraphics[width=2in,height=4in,angle=-90]{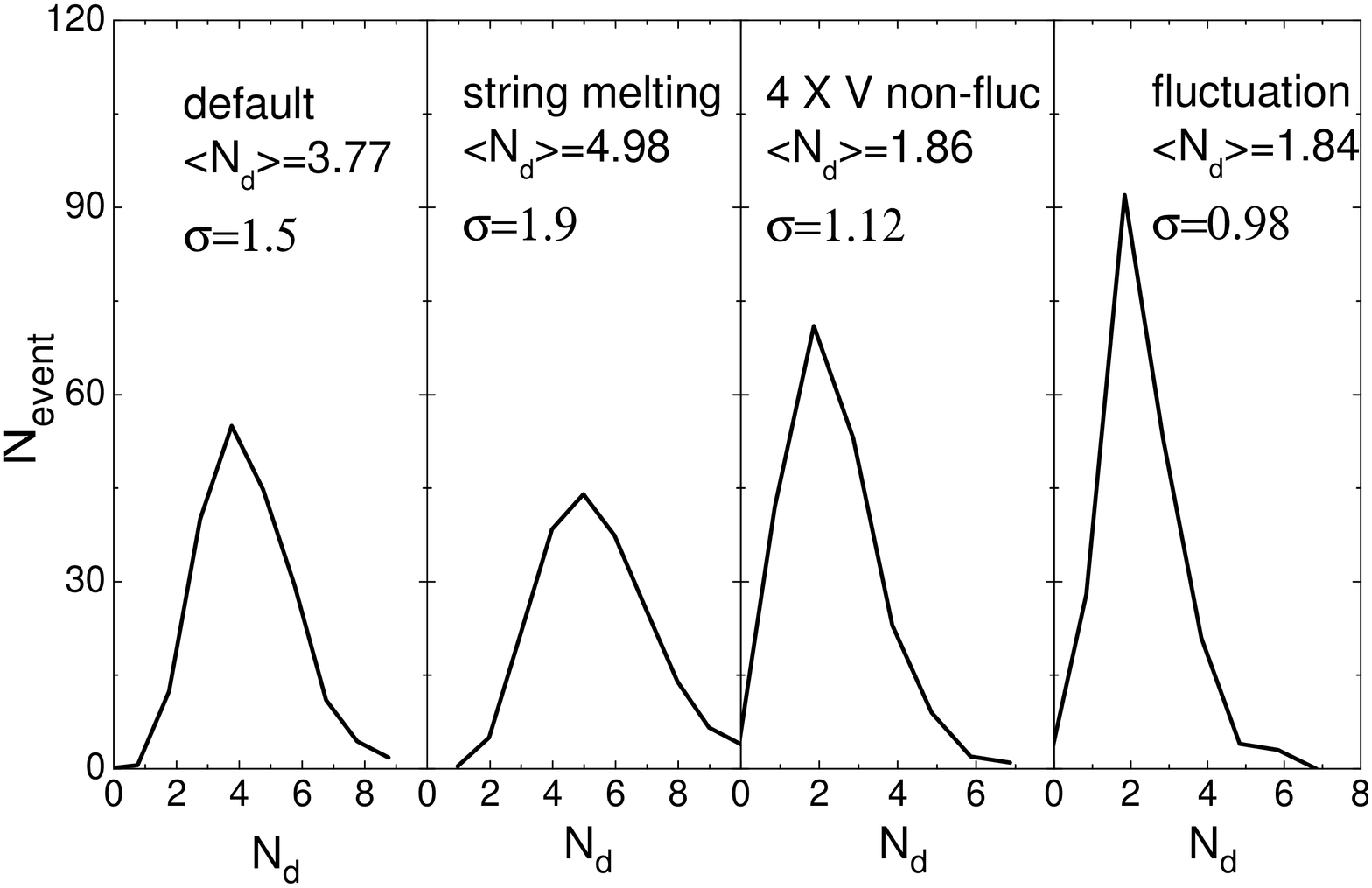}}
\caption{(Color online) Top window: Initial (top panel) and final
(lower panel) hadronic spatial distributions in different scenarios.
Middle window: Fluctuation of hadron mean transverse momentum.
Bottom window: Fluctuation of final deuteron
number. All are for central Au+Au collisions at $\sqrt {s_{NN}}=7$ GeV.}
\label{fluctuation}
\end{figure}

As mentioned previously, the baryon-rich partonic matter produced in
heavy ion collisions in RHIC low-energy runs and at FAIR is expected
to undergo a first-order phase transition to the hadronic matter,
instead of the smooth crossover transition of the quark-gluon plasma
with low baryon chemical potential that is produced in RHIC high
energy runs. To model the effect due to a first-order quark-gluon
plasma to hadronic matter phase transition, which is at present
absent in the AMPT model, a large density fluctuation is introduced
in the hadronic matter formed immediately after hadronization.
Specifically, we assume that the hadronic matter after a first-order
phase transition consists of clusters of various sizes. This is
achieved by redistributing hadrons produced in the AMPT with string
melting in a volume that is four times larger but keeping their
momenta unchanged. Furthermore, the average number of clusters is
taken to be five and follows a Poisson distribution as in the
analysis of the nuclear matter gas-liquid phase transition in
low-energy heavy ion collisions \cite{randrup}. An example of the
hadron spatial distribution is shown in the top right panel of the
top window in Fig.~\ref{fluctuation}, and it is compared with those
in the default AMPT, AMPT with string melting, and AMPT with string
melting together with four times larger initial hadronic matter
volume but without fluctuation, shown in other panels of the top
window. The final hadronic spatial distributions in these four
scenarios are shown in the lower panels of the top window and are
seen to be different. As probes of these different scenarios, we
have considered the fluctuation of final hadron mean transverse
momentum (middle window) as well as that of final deuteron yield
(bottom window), which is obtained by the coalescence model using
the proton and neutron phase distributions at freeze out
\cite{chenthe3}. It is seen that the root mean square fluctuation of
final hadron mean transverse momentum is smaller than that of
initial hadron mean transverse momentum, and its value is similar in
all scenarios. As a result, the mean transverse momentum fluctuation
is not sensitive to the initial density fluctuation of produced
hadrons. For the deuteron number fluctuation, increasing the initial
volume of the hadronic matter reduces the final deuteron yield
compared to those from the default AMPT and AMPT with string
melting. The final deuteron number is, however, essentially
independent of whether there is an initial fluctuation in the
hadronic matter density. Although the initial density fluctuation is
destroyed by final-state hadronic scatterings, it cannot exist
without an increasing initial volume of hadronic matter as a result
of first-order phase transition.

\section{Two-pion correlations}

Particle interferometry based on the Hanbury-Brown Twiss (HBT)
effect can provide information not only on the spatial extent of the
emission source but also on its expansion velocity and emission
duration \cite{Pratt:su,bertsch,Bertsch:1988db,Pratt:zq}. In
particular, the long emission time as a result of phase transition
from the quark-gluon plasma to the hadronic matter in relativistic
heavy ion collisions is expected to lead to an emission source which
has a much larger radius in the direction of the total transverse
momentum of detected two particles ($R_{\rm out}$) than that
($R_{\rm side}$) perpendicular to both this direction and the beam
direction ($R_{\rm long}$). Although the extracted ratio $R_{\rm
out}/R_{\rm side}$ from a Gaussian fit to the measured two-pion
correlation function in Au+Au collisions at $\sqrt{s_{NN}}=130$ GeV
is close to one \cite{Adler:2001zd,Johnson:2001zi,Adcox:2002uc}, the
source function extracted from the imaging method seems to show a
longer tail in the out direction compared to other directions
\cite{Afanasiev:2007kk}. The small value of $R_{\rm out}/R_{\rm
side}$ has been attributed to strong space-time and momentum
correlations in the emission source \cite{Tomasik:1998qt}. Since the
quark-gluon plasma produced in RHIC low-energy runs and at FAIR is
expected to undergo a first-order phase transition, the emission
source is expected to also have a large radius in the out direction.

\begin{figure}[ht]
\vspace{-0.2cm}\centerline{
\includegraphics[width=3in,height=3in,angle=0]{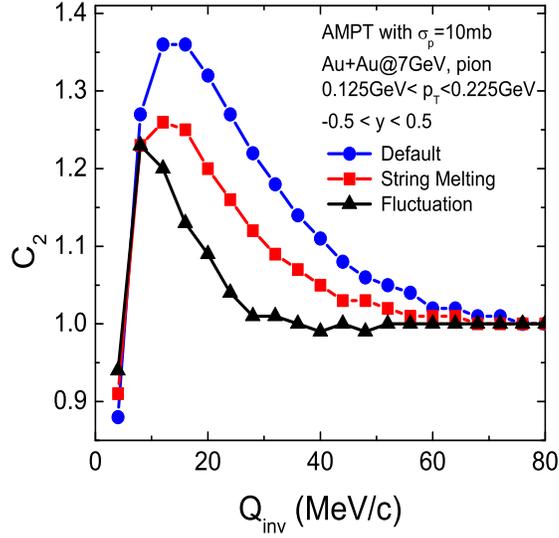}}
\caption{(Color online) Two-pion correlation functions in central
Au+Au collisions at $\sqrt {s_{NN}}=7$ GeV.}\label{hbt}
\end{figure}

Using the emission function obtained from the AMPT model with string
melting and a parton scattering cross section of $10$ mb for central
($b=0$ fm) Au+Au collisions at $\sqrt{s_{NN}}=7$ GeV, we have
evaluated the correlation function $C_2({\bf Q},{\bf K})$ in the
longitudinally comoving frame using the program Correlation After
Burner \cite{pratt:uf} that takes into account final-state strong
and Coulomb interactions between two charged pions. In Fig.
\ref{hbt}, we show the calculated correlation function including
final-state Coulomb interactions for midrapidity ($-0.5<y<0.5$)
charged pions with transverse momentum $125<p_{\rm T}<225$ MeV$/c$
as a function of invariant momentum in the three scenarios of
default AMPT (filled circles), AMPT with string melting (filled
squares), and AMPT with density fluctuation after hadronization
(filled triangles). It is seen that the correlation function becomes
narrower when a partonic matter is formed, and the width is further
reduced if the initial volume of the hadronic matter after phase
transition becomes larger as a result of the first-order phase
transition.

\section{Charm suppression}

The partonic matter with a finite baryon chemical potential that is
expected to be formed in RHIC low energy runs and at FAIR offers a
possibility to test the idea recently introduced in Ref.\cite{hees}
that the large charm quark elliptic flow observed in high energy runs at RHIC
is a result of resonance scattering between charm
quark and anti-light quarks in the produced partonic matter.
Specifically, the scattering cross section between a charm quark and
an light antiquark or a charm antiquark and a light quark is given
by
\begin{eqnarray}\label{resonance}
\sigma_{{\bar c}q\to{\bar c}q}(s^{1/2}) =\frac{1}{9}\frac{2J+1}{4}\frac{\pi}{k^2}
\frac{\Gamma_D^2}{(s^{1/2}-m_D)^2+\Gamma_D^2/4}
\end{eqnarray}
in terms of the charm meson mass $m_D$, width $\Gamma_D$ and spin
$J$. Taking $m_D=2~{\rm GeV}$, $\Gamma_D=0.3-0.5~{\rm GeV}$, charm
quark mass $m_c=1.5~{\rm GeV}$, light quark mass $m_q=5-10~{\rm
MeV}$ and including scalar, pseudoscalar, vector and axial vector
charmed mesons gives a peak cross section of about 6 mb at the
charmed meson mass. The drag coefficient for charm quark or
antiquark in a baryon-free quark-gluon plasma can then be calculated
using the scattering amplitude ${\cal M}$ of the resonance
scattering via
\begin{equation} \gamma(|{\bf
p}|,T)=\sum_i\left(\langle\overline{|{\cal M}|^2}\rangle
-\langle\overline{|{\cal M}|^2}{\bf p}\cdot{\bf p^\prime}\rangle
/|\bf p|^2\right).
\end{equation}
In the above, ${\bf p}$ and ${\bf p}^\prime$ are, respectively, the
momenta of the heavy quark before and after a collision. Since the
resonance scattering cross section is isotropic, it leads to a much
larger drag coefficient of $\gamma\sim 0.16 c/{\rm fm}$ than that
given by the pQCD $t$-channel diagrams. Since there are about equal
numbers of light quark and antiquarks in the quark-gluon plasma
produced in high energy runs at RHIC, charm and anticharks thus
lose same appreciable energy when traversing through the baryon-free
quark-gluon plasma.

\begin{figure}[ht]
\centerline{
\includegraphics[width=3in,height=3in,angle=-90]{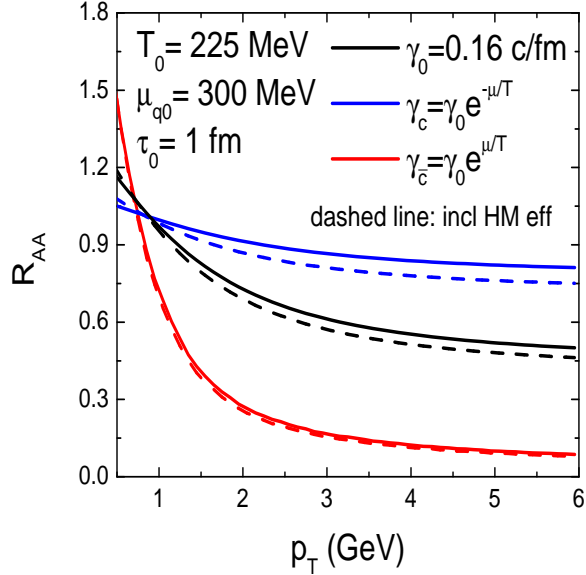}}
\caption{(Color online) Nuclear suppression factor for charmed meson
and anticharmed mesons in baryon-rich quark-gluon
plasma.}\label{charm}
\end{figure}

For heavy ion collisions in RHIC low-energy runs and at FAIR,
resonance scattering affects charm and anticharm quarks differently
as there are more light quarks than light antiquarks in the
baryon-rich quark-gluon plasma produced in these collisions, with
the quark drag coefficients decreased by the factor $e^{-\mu/T}$ for
charm quarks and increased by the factor $e^{\mu/T}$ for anti-charm
quarks. For Au+Au collisions at $E_{\rm beam}=40$ AGeV, the initial
temperature and baryon chemical potential of produced quark-gluon
plasma are then $T_0=225$ MeV and $\mu_0=300$ MeV according to the
AMPT model as shown in Fig.\ref{phase}. Taking the quark-gluon
plasma to have a spherical shape with an initial radius $R_0=7$ fm,
the time evolution of the radius can then be determined from the
time evolution of temperature and baryon chemical potential given in
Fig.\ref{phase} if we assume that the entropy of the quark-gluon
plasma is conserved during its expansion. For the distribution of
initially produced charm and anti-charm quarks, their momentum
spectra are generated by the PYTHIA program \cite{pythia} and they
are assumed to be produced uniformly inside the quark-gluon plasma.
As charm and anti-charm quarks traverse through the expanding
quark-gluon plasma, they loose their momentum according to
$dp/dt\approx -\gamma p$. The resulting nuclear modification factors
$R_{AA}$ for charm and anti-charm quarks, defined as the ratios of
their final to initial momentum spectra are shown in
Fig.\ref{charm}. It is seen that the nuclear modification factor for
anti-charm quarks is much smaller than that for charm quarks. This
result is not much affected by final-state hadronic reactions as
also shown in the figure. We therefore expect that the momentum
spectrum of charmed mesons will be less quenched than that of
anti-charmed mesons in heavy ion collisions in RHIC low-energy runs
and at FAIR, which thus provides a very interesting opportunity to
study the dynamics of charm and anti-charm quarks in the quark-gluon
plasma.

\section{Seeing QCD first-order phase transition with phi mesons}

\begin{figure}[ht]
\vspace{1.9cm}
\centerline{
\includegraphics[width=3in,height=3in,angle=-90]{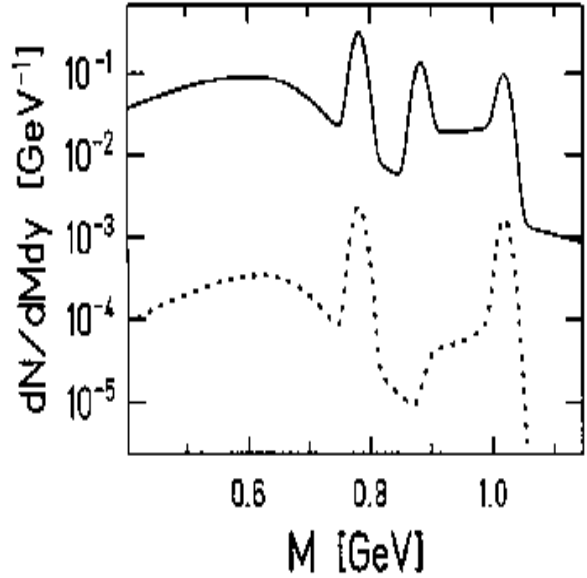}}
\caption{The dilepton invariant mass spectrum in heavy ion
collisions. The solid line is obtained with a first-order phase
transition while the dotted line is from the scenario without the quark-gluon plasma.
From Ref.\cite{asakawa94a}.}\label{phi}
\end{figure}

Another possible way to verify the existence of a first-order QGP to
hadronic matter phase transition in heavy ion collisions is to study
phi meson production through its emitted dileptons. Because of its
narrow width of $\sim 4~{\rm MeV}$, a phi meson in heavy ion
collisions normally decays to dileptons mainly at the time when it
freezes out from the hadronic matter. If there is a first-order QGP
to hadronic matter phase transition, dileptons from phi mesons decay
during the mixed phase of the first-order phase transition can also
be appreciable as a result of the long duration of the mixed phase.
Since the phi meson mass may be reduced in hot and dense medium, the
emitted dileptons from phi mesons in the mixed phase thus have lower
invariant masses than those from phi mesons decay at freeze out. As
a result, the dilepton invariant mass spectra would have an
additional peak at a lower mass than that corresponding to the free
phi mass mass. In Ref.~\cite{asakawa94a}, this effect is studied in
a hydrodynamic model for heavy ion collisions that includes a
first-order phase transition. With a temperature-dependent phi meson
mass determined from the QCD sum rule \cite{asakawa94b}, the
resulting phi meson invariance spectrum indeed shows another peak
between the $\rho/\omega$ and free phi meson peaks as shown by the
solid line in Fig.~\ref{phi}.  Although the double phi peak feature
in the dilepton invariant mass spectrum would remain if the QGP to
hadronic matter phase transition is a crossover but lasts a
sufficient long time \cite{asakawa94c}, it becomes a shoulder in the
absence of the QGP phase as shown by the dashed line in the figure.
Since the flow velocity is smaller at phase transition than at
hadronic freeze out, the inverse slope parameter of the low mass phi
meson transverse momentum spectrum is also expected to be smaller
than that of the transverse momentum spectrum of normal phi mesons
and thus to reflect more closely the phase transition temperature.

\section{Conclusions}

Heavy ion collisions in the low-energy runs at RHIC and at FAIR is
expected to produce a quark-gluon plasma that has a finite baryon
chemical potential, allowing thus the possibility to study the
location of the critical endpoint in the QCD phase diagram, where
the quark-gluon plasma to hadronic matter transition changes from a
crossover to a first-order phase transition. Using a multiphase
transport model, that includes interactions in both initial partonic
and final hadronic matters and the transition between these two
phases of matter via a quark coalescence model, we have studied the
effects of partonic interactions on the elliptic flow, two-pion
correlations as well as the effect of density fluctuations due to a
first-order transition between the quark-gluon plasma and hadronic
matter on fluctuations of hadron mean transverse momentum and
produced deuteron number. We have found that strong partonic
interactions during the partonic stage of the collisions enhance
significantly the particle elliptic flow and also leads to an
approximate constituent quark number scaling of the elliptic flows
of identified hadrons. Although the presence of density fluctuation
due to a first-order phase transition has little effect on the
fluctuation of final hadron mean transverse momentum, it affects
appreciably the yield of produced deuterons and the width of the
two-pion correlation function.  We have further found that the net
baryon-rich quark-gluon plasma formed in these collisions provides
the possibility to test the resonance scattering mechanism of charm
energy loss in quark-gluon plasma, which has been quite successful
in understanding the charm energy loss and flow in heavy ion
collisions at RHIC, i.e., a larger energy loss of anti-charm quarks
than quarks. The long duration expected for a first-order phase
transition can also lead to the appearance of a lower mass peak
between the $\rho/\omega$ and $\phi$ mesons in the dilepton
invariance mass spectrum as a result of the lower phi meson mass in
the mixed phase of the quark-gluon plasma to hadronic matter
transition.  Many interesting phenomena are thus expected to be
observed in RHIC low-energy runs and at FAIR, and they not only are
of interest in their own right but will also provide the possibility
to study the location of the critical endpoint in the QCD phase
diagram, where the crossover transition at low baryon chemical
potential changes to a first-order transition when the baryon
chemical becomes sufficient large.

\section*{Acknowledgements}

This talk was based on work supported by the US National Science
Foundation under Grant Nos. PHY-0758115 and the Welch Foundation
under Grant No. A-1358 (C.M.K.), the NNSF of China under Grant Nos.
10405011 (BWZ), 10575071 and 10675082, MOE of China under project
NCET-05-0392, Shanghai Rising-Star Program under Grant No.
06QA14024, and the SRF for ROCS, SEM of China (L.W.C.).

\end{document}